\documentclass[twocolumn,dvipsnames]{aastex62}
\pdfoutput=1 
\usepackage{amsmath,amstext}
\usepackage{apjfonts}
\usepackage[figure,figure*]{hypcap}
\usepackage{ulem}
\usepackage{xspace}

\usepackage{graphicx}
\usepackage{natbib}
\usepackage{amssymb}
\usepackage{amsmath}
\usepackage{array,booktabs}
\usepackage{mathptmx}
\usepackage{booktabs}
\usepackage{hyperref}
\usepackage{float}

\DeclareSymbolFont{cmletters}{OML}{cmm}{m}{it}
\DeclareMathSymbol{v}{\mathalpha}{cmletters}{"76}

\newcommand{\msun}{\,{M_{\odot}}}
\newcommand{\cm}{\,{\rm cm}}
\newcommand{\s}{\,{\rm s}}	

\newcommand{\vmin}{\,{v_{\rm min}}}
\newcommand{\vesc}{\,{v_{\rm esc}}}
\newcommand{\vmax}{\,{v_{\rm max}}}

\newcommand{\rg}{\,{r_{\rm g}}}
\newcommand{\Mej}{\,{M_{\rm ej}}}

\newcommand{\appropto}{\mathrel{\vcenter{
			\offinterlineskip\halign{\hfil$##$\cr
				\propto\cr\noalign{\kern2pt}\sim\cr\noalign{\kern-2pt}}}}}
\usepackage{xcolor}

\shorttitle{sGRB jet-ejecta interaction in GRMHD simulations}
\shortauthors{Gottlieb et al.}

\begin{document}
	\title{On the jet-ejecta interaction in 3D GRMHD simulations of a binary neutron star merger aftermath}
	
	\author[0000-0003-3115-2456]{Ore Gottlieb}
	\email{ore@northwestern.edu}
	\affiliation{Center for Interdisciplinary Exploration \& Research in Astrophysics (CIERA), Physics \& Astronomy, Northwestern University, Evanston, IL 60202, USA}
	
	\author[0000-0001-8200-5769]{Serena Moseley}
	\affiliation{Center for Interdisciplinary Exploration \& Research in Astrophysics (CIERA), Physics \& Astronomy, Northwestern University, Evanston, IL 60202, USA}
	
	\author[0000-0003-0994-115X]{Teresita Ramirez-Aguilar}
	\affiliation{Center for Interdisciplinary Exploration \& Research in Astrophysics (CIERA), Physics \& Astronomy, Northwestern University, Evanston, IL 60202, USA}

	\author[0000-0003-2333-6116]{Ariadna Murguia-Berthier}
	\affiliation{Center for Interdisciplinary Exploration \& Research in Astrophysics (CIERA), Physics \& Astronomy, Northwestern University, Evanston, IL 60202, USA}
	
    \author{Matthew Liska}
    \affiliation{Institute for Theory and Computation, Harvard University, 60 Garden Street, Cambridge, MA 02138, USA; John Harvard Distinguished Science and ITC}
	
	\author[0000-0002-9182-2047]{Alexander Tchekhovskoy}
	\affiliation{Center for Interdisciplinary Exploration \& Research in Astrophysics (CIERA), Physics \& Astronomy, Northwestern University, Evanston, IL 60202, USA}
	
\begin{abstract}
		
Short $\gamma$-ray burst (sGRB) jets form in the aftermath of a neutron star merger, drill through disk winds and dynamical ejecta, and extend over four to five orders of magnitude in distance before breaking out of the ejecta. We present the first 3D general-relativistic magnetohydrodynamic sGRB simulations to span this enormous scale separation. They feature three possible outcomes: jet+cocoon, cocoon, and neither. Typical sGRB jets break out of the dynamical ejecta if (i) the {\it bound} ejecta's isotropic equivalent mass along the pole at the time of the BH formation is $ \lesssim10^{-4}~{\rm M_{\odot}} $, setting a limit on the delay time between the merger and BH formation, otherwise, the jets perish inside the ejecta and leave the jet-inflated cocoon to power a low-luminosity sGRB; (ii) the post-merger remnant disk contains strong large-scale vertical magnetic field, $\gtrsim10^{15}$ G; and (iii) if the jets are weak ($\lesssim10^{50}$~erg), the ejecta's isotropic equivalent mass along the pole must be small ($\lesssim10^{-2}~{\rm M_{\odot}}$). Generally, the jet structure is shaped by the early interaction with disk winds rather than the dynamical ejecta. As long as our jets break out of the ejecta, they retain a significant magnetization ($\lesssim1$), suggesting that magnetic reconnection is a fundamental property of sGRB emission. The angular structure of the outflow isotropic equivalent energy after breakout consistently features a flat core followed by a steep power-law distribution  (slope $\gtrsim3$), similar to hydrodynamic jets. In the cocoon-only outcome, the dynamical ejecta broadens the outflow angular distribution and flattens it (slope $\sim1.5$).
		
\end{abstract}
	
	\section{Introduction}\label{sec:introduction}
	
	The discovery of the first binary neutron star (BNS) merger, GW170817, marked the beginning of the multi-messenger era with the simultaneous detections of gravitational waves (GW) and electromagnetic (EM) waves across the entire electromagnetic spectrum \citep[see reviews in][]{Nakar2019,Margutti2021}.
	The multi-band observations spanned $8$ orders of magnitude in time and brought to light three key components in the merger aftermath:
	(i)~the dynamical ejecta, which was stripped from the neutron stars (NSs) during the merger, ultimately produced the optical-infrared $ \sim $week long kilonova signal powered by the $ \beta $-decay of $ \Mej \sim 5\times 10^{-2}\msun $ heavy $ r $-process elements \citep[e.g.,][]{Kasen2017,Metzger2017};
	(ii)~the relativistic short $ \gamma $-ray burst (sGRB) jet, whose emission was at first beamed away from Earth at a viewing angle $ \theta_{\rm obs} \approx 20^\circ $ with respect to the jet axis, was revealed $ \sim 100 $ days after the merger, thanks to its radio/X-ray afterglow emission \citep{Mooley2018}; and
	(iii)~the mildly-relativistic cocoon, which was generated as the jet propagated through the dense medium surrounding the merger site. The cocoon powered the first $ \sim 100 $ days of the radio/X-ray afterglow signal before the jet became visible, and is also a leading candidate for the source of the $ \gamma $-rays 1.7s after the merger via shock breakout emission \citep{Gottlieb2018b}.
	
	While significant progress in our understanding of the post-merger dynamics has been achieved following GW170817, the early evolution of the system while it is still optically thick cannot be addressed directly through observations. Inferring the system properties using analytic models of emission (such as the afterglow) is also difficult because those include multiple degeneracies that limit the information that can be extracted from observations \citep[e.g.,][]{Nakar2021}. Numerical simulations are thus an essential tool to study the interaction between the three aforementioned components and its effect on the emerging outflow structure and emission.
	
	The understanding of the jet-ejecta interaction has been considerably improved thanks to a wide range of numerical studies in the past years  \citep{Kasliwal2017,Lazzati2017,Duffell2018,Gottlieb2018a,Gottlieb2018b,Gottlieb2020b,Gottlieb2021a,Kathirgamaraju2018,Geng2019,Lazzati2019,Gottlieb2020c,Ito2021,Klion2021,Murguia-Berthier2021,Pavan2021,Urrutia2021,Lamb2022,Nativi2022}. However, these studies have been subject to two major limitations:
	
	(i) They ignored the dominant magnetic energy contribution to the jets (some considered subdominant magnetic fields).
	\citet[][hereafter GN21]{Gottlieb2021d} conducted a comprehensive analytic and numerical study of the propagation of hydrodynamic and weakly magnetized jets through the ejecta. Their analysis showed that when the jet head moves sub-relativistically, an unmagnetized jet propagates $2{-}3$ times slower than a jet with a toroidal magnetic field of magnetization $ \sigma  \equiv b^2/(4\pi\rho c^2) \sim 10^{-2}{-}10^{-1} $, where $ b $ and $ \rho $ are the comoving magnetic field strength and mass density, respectively. They further found that such toroidal magnetic fields can shorten the jet breakout time from the ejecta:  by up to an order of magnitude if the breakout time is longer than the delay time between the merger and the jet launching, and by a factor of two otherwise.
	Overall, magnetic fields, which appear to be essential for the jet launching from a compact object \citep[e.g.][]{Kawanaka2013,Leng2014}, were shown to also play a central role in the evolution and breakout of the jet from the ejecta. However, the propagation of highly magnetized sGRB jets has not been possible until now.
	
	(ii) They injected the jet from the grid boundary rather than letting it form self-consistently as part of general relativistic (GR) magnetohydrodynamic (MHD) processes in the merger, thereby prescribing its intrinsic properties such as the opening angle, power, structure, etc.
	
	Numerical simulations in which Poynting-flux dominated jets emerge from the compact object naturally avoid the need to prescribe the jet properties and can therefore address both of the above-mentioned limitations.
	The first 3D GRMHD studies of jets in NS mergers used a post-merger initial setup, which consisted of a compact object, a black hole (BH), and an accretion disk, but omitted the dynamical ejecta \citep{Christie2019,Fernandez2019,Kathirgamaraju2019}. 
	
	The first attempt to model the interaction between a self-consistent launched jet and the ejecta in a 3D GRMHD setup was done by \citet{Nathanail2021}. Their initial setup results in a jet breakout from the ejecta at $ r \approx 10^8 $ cm, just a few milliseconds after the merger, implying that the role of the ejecta was negligible. Furthermore, the $ \gamma $-ray signal in GW170817, which was generated 1.7 s after the merger, suggests that the breakout took place on $ \sim 1 $ s timescale. Recently, \citet{Gottlieb2021c} performed a 3D GRMHD simulation of a self-consistent jet launching with an initial setup motivated by GW170817 observables, but their simulation lasted only $ \sim 0.4 $ s, before the jet managed to break free from the ejecta and fully form the outflow.
	
	Using a novel adaptive mesh refinement (AMR) technique, here we perform the suite of the highest-resolution 3D GRMHD simulations of sGRB jets to date. This enables us to study how the physical parameters of the merger affect the resulting outflow close to the emission zone. In \S\ref{sec:setup}, we describe the physical and numerical setup of the simulations. In \S\ref{sec:launching}, we present the conditions at the jet launching site, and in \S\ref{sec:propagation} we discuss the jet propagation through the ejecta and its post-breakout structure and properties. In \S\ref{sec:conclusions}, we conclude.
	
	\section{Numerical setup}\label{sec:setup}
	
	We assume that the central BH forms with a short time delay, $t_d$, after the merger, e.g., after a brief phase of an unstable hypermassive NS that collapses to form the BH. The delay time can be constrained from the ejecta mass and composition, as well as the jet propagation. Estimates of $ t_d $ place it in the range of $ 0.1\s \lesssim t_d \lesssim 1\s $ \citep[e.g.,][]{Metzger2018,Nakar2019,Murguia-Berthier2021}. The BH mass and dimensionless spin are set to be $ M_{\rm BH} = 3\msun $ and $ a = 0.9375 $, respectively. The BH is surrounded by a torus, which formed during the merger. We choose the torus solution following \citet{FM1976}, and set its inner radius to $r_{\rm in} = 6\rg $, the radius of maximum pressure to $r_{\rm max} = 12\rg $, and its mass to $ M_t = 4 \Mej $, as suggested by numerical simulations of BNS merger remnants \citep{Radice2016,Siegel2017,Christie2019,Fernandez2019}. Into the torus we insert a large-scale magnetic field with plasma beta parameter $ \beta_p \equiv p_g/p_m $, where $ p_g $ and $ p_m $ are the gas and magnetic pressure, respectively. The magnetic vector potential depends on the field geometry. We explore two configurations: toroidal,
	\begin{equation}
		\textbf{A} = A_{\hat{\theta}}(r,\theta)\hat{\theta}=B_0\rg\max\left(\frac{\rho(r,\theta)}{\rho_{\rm max}}-5\times 10^{-4}, 0 \right)\hat{\theta},
	\end{equation}
	and poloidal,
	\begin{equation}
		\textbf{A} = A_{\hat{\phi}}(r,\theta)\hat{\phi}=\frac{B_0r^2}{\rg}\max\left(\frac{\rho(r,\theta)}{\rho_{\rm max}}-5\times 10^{-4}, 0 \right)\hat{\phi},
	\end{equation}
	where $ B_0 $ is the characteristic magnetic field strength in the torus determined by $ \beta_p $, $ \rg $ is the BH gravitational radius, and $ \rho_{\rm max} $ is the maximum density in the torus.


	\begin{table}
		\setlength{\tabcolsep}{2.4pt}
		\centering
		\renewcommand{\arraystretch}{1.3}
		\begin{tabular}{c|c c c c c c c c c}
			
			Model & $ \Mej~[\msun] $ & $ \alpha $ & $ \delta $ & $ t_d~[\s] $ & $ \frac{\vmin}{c} $ & $ \textbf{A} $ & $ \beta_p $ & $ t_b~[{\rm s}] $ & $ E_c~[{\rm erg}] $
			\\	\hline
			$ \alpha $3d5 & 0.05 & 3 & 1 & 0.5 & 0.05 & $ A_\phi $ & $ 10^3 $ & 0.37 & $ 8\times 10^{49} $ \\
			$ \alpha $1d5 & 0.05 & 1 & 1 & 0.5 & 0.05 & $ A_\phi $ & $ 10^3 $ & 0.34 & $ 5\times 10^{49} $ \\
			$ \alpha $3d1 & 0.05 & 3 & 1 & 0.1 & 0.07 & $ A_\phi $ & $ 10^3 $ & 0.16 & $ 3\times 10^{49} $ \\
			$ \alpha $3d1Mb & 0.05 & 3 & 1 & 0.1 & 0.05 & $ A_\phi $ & $ 10^3 $ & 0.16 & $ 3\times 10^{49} $ \\
			$ \alpha $3d1MbBs & 0.05 & 3 & 1 & 0.1 & 0.05 & $ A_\phi $ & $ 10 $ & 0.16 & $ 3\times 10^{51} $\\
			$ \alpha $3d5iso & 0.05 & 3 & 0 & 0.5 & 0.05 & $ A_\phi $ & $ 10^3 $ & 0.50 & $ 1\times 10^{50} $ \\
			$ \alpha $3d5Bw & 0.05 & 3 & 1 & 0.5 & 0.05 & $ A_\phi $ & $ 10^4 $ & 0.76 & $ 8\times 10^{48} $ \\
			$ V $Bw & 0.00 & - & - & - & - & $ A_\phi $ & $ 10^4 $ & - & - \\
            $ \alpha $3d5Bwiso & 0.05 & 3 & 0 & 0.5 & 0.05 & $ A_\phi $ & $ 10^5 $ & $ \infty $ & $ 4\times 10^{47} $ \\
			$ \alpha $3d5Bt & 0.05 & 3 & 1 & 0.5 & 0.05 & $ A_\theta $ & $ 1 $ & $ \infty $ & $ 5\times 10^{47} $ \\
			$ V $ & 0.00 & - & - & - & - & $ A_\phi $ & $ 10^3 $ & - & - \\
		\end{tabular}
		
		\caption{
			A summary of the models' parameters. The model names describe the $\alpha$ and $t_{d}$ for each model. The subscripts of B indicate a strong, weak, or toroidal magnetic field, Mb indicates that the dynamical ejecta contains bound mass due to $ \vmin < \vesc $, $ V $ denotes no dynamical ejecta. $ \Mej $ is the total ejecta mass, $ \alpha $ is the radial ejecta mass density power-law index, $ \delta $ signifies the inclusion or exclusion of an angular component in the ejecta density profile, $ t_d $ is the time delay between the merger and the formation of the BH, $ \textbf{A} $ is the magnetic vector potential, which can be either poloidal ($ A_\phi $) or toroidal ($ A_\theta $), $ \beta_p $ is the minimal plasma beta parameter in the torus, $ t_b $ is the average breakout time (between the two jets) with respect to the time of the BH formation, and $ E_c $ is the energy in the cocoon, estimated as the jet energy that reaches the jet head before breakout, $ E_c \approx \int_0^{t_b-\chi}{L_{j,0}dt} $, where $ \chi \equiv \frac{\vmax}{c}\left(t_b+t_d\right) $ to account for the jet material that does not cross the reverse shock by the time the jet head breaks out.
			Model $ V $Bw is identical to model $ \alpha $3d5Bw, but without any ejecta, and model $ \alpha $3d1 is identical to model $ \alpha $3d1Mb, but with $ \vmin > \vesc $. Both of these models are carried out to examine whether the jet is successfully launched, and thus are not modeled at late times.
    		}
    		\label{tab:models}
	\end{table}
	
	The initial distribution of the dynamical ejecta is motivated by GW170817 observations. We assume unmagnetized cold dynamical ejecta of mass $ \Mej = 0.05 \msun $ that expands homologously with the maximum velocity inferred from the GW170817 kilonova, $ \vmax = 0.25 c$ \citep{Kasen2017}. The minimum ejecta velocity is obtained via the escape velocity of the ejecta from the BH, $ \vmin \approx \vesc = (2GM_{\rm BH}/t_d)^{1/3} \approx 0.05 $ c. Due to the uncertainties in the ejecta density distribution, we adopt a mass density profile with both radial and angular distributions as follows\footnote{In nature the ejecta likely includes anisotropies that may alter the jet evolution \citep{Pavan2021}. This will be addressed in a follow-up paper where the ejecta will be set self-consistently by remapping it from numerical relativity simulations.}. The radial part is modeled by a power-law distribution $ r^{-\alpha} $. If the ejecta is isotropic, then the comparable amount of mass in the slow and fast components of the GW170817 ejecta implies that the radial power-law index is $\alpha = 3 $, which we adopt as the canonical value. We adopt the angular distribution by fitting numerical simulations of \citet{Nedora2021}, which we approximate via an order of magnitude density ratio between the polar and equatorial components, with the angular dependence of $ \sin^2\theta $ in between. This results in the following homologous ejecta mass density profile,
	\begin{equation}\label{eq:density}
		\rho(\vmin t_d<r<\vmax t_d,\theta) = \rho_0r^{-\alpha}\left(0.1+{\rm sin}^2\theta\right)^\delta~.
	\end{equation}
	where $ \rho_0 $ is set by the requirement of $ \Mej = 0.05\msun $, and $ \delta $ determines whether the ejecta has an angular component or is isotropic.
	We carry out a set of simulations with different values of $ \alpha, \delta, t_d, \vmin, \beta_p $ and $ \textbf{A} $, two simulations without dynamical ejecta, and two simulations with an initial bound mass in the ejecta, as listed in Table~\ref{tab:models}. Of these, we designate a fiducial model, $\alpha$3d5, that starts with poloidal magnetic flux geometry and typical value of the plasma beta parameter, $\min\beta_p = 10^3$ in the torus. It also adopts fiducial values for the mass, $M_{\rm ej} = 0.05\msun$, radial density power-law index, $\alpha = 3$, and time delay, $t_d = 0.5$~s for the dynamical ejecta.
	
	We perform the simulations using our 3D GPU-accelerated code \textsc{h-amr} \citep{Liska2019}, where we employ an ideal equation of state with the adiabatic index of $ 4/3 $, as appropriate for relativistic, radiation dominated gas. For numerical stability purposes we set a density floor in the code by setting the maximum magnetization in the simulation to $ \sigma_0 = 20 $, which is roughly the maximum asymptotic Lorentz factor to which the jets can accelerate. 
	
	We carry out the simulations in the spherical polar coordinates, $r$, $\theta$, $\phi$. We tilt the entire simulation (both the BH metric and the initial conditions) by $ 90^\circ $ in order to ensure the free development of 3D instabilities in the jets by directing them away from the polar axis. However, in order to avoid confusion, we will still refer to the $ \hat{z} $-axis as the BH rotation axis and $ \theta $ as the angle relative to that axis. 
	
	The simulations utilize local adaptive time-stepping and AMR techniques. The refinement criterion identifies the jet and cocoon regions via pre-set ranges of magnetization values. If either one of these regions contains fewer than the desired number of cells -- 96 cells in the $\theta$- or 192 cells in the $\phi$-directions -- then the region is refined, until the desired number of cells is reached.
	Our spherical grid uses a logarithmic cell distribution along the $r$-direction, ranging from just inside the event horizon ($4.5\times 10^5\cm$) out to $4.5\times 10^{10}\cm$, and uniform cell distribution along $\theta$- and $\phi$-directions. The base grid contains $N_r\times N_\theta \times N_\phi = 384 \times 96 \times 192$ cells in the $r$-, $\theta$- and $\phi$-directions, respectively. We use up to 3 levels of AMR. As a result, at the maximum refinement level, the grid resolution is $N_r\times N_\theta \times N_\phi = 3072 \times 768 \times 1536$.
	We verify that the above resolution is sufficient to resolve the wavelength of the fastest growing magnetorotational instability (MRI) mode \citep{1991ApJ...376..214B}. We do this by calculating the MRI quality factor $Q$, which gives the number of cells per the MRI wavelength. We find that $ Q \gtrsim 100 $ (in both the $ \hat{\theta} $ and $ \hat{\phi} $ directions) at all times in all models, much larger than the $ Q \sim 10 $ that is required to properly resolve MRI \citep[e.g.,][]{Hawley2011}, except for model $ \alpha $3d5Bwiso in the first few milliseconds, during which our solution is not valid.
	
	\section{Jet launching}\label{sec:launching}
	
	\begin{figure}
		\centering
		\includegraphics[scale=0.22,trim=0 0 0 0]{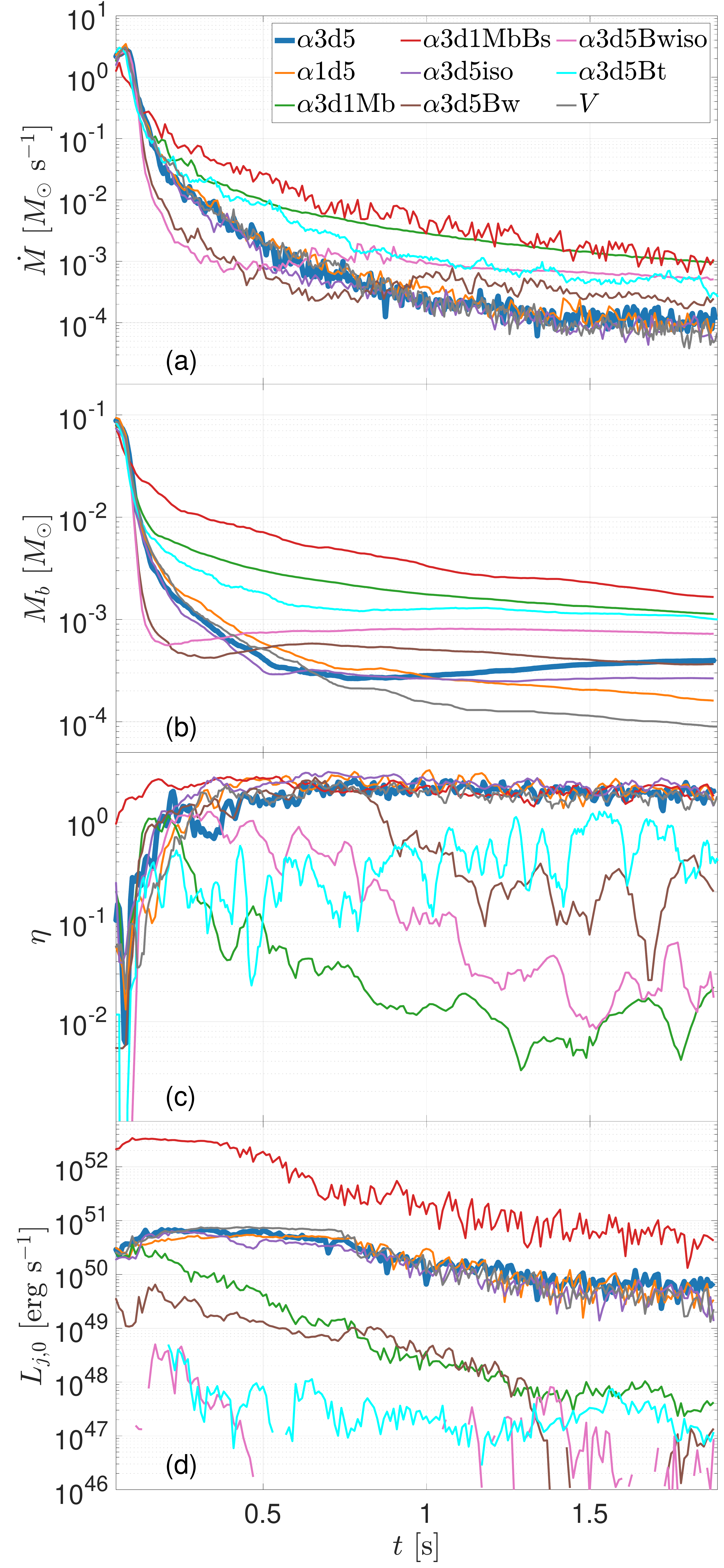}
		\caption[]{
			Time evolution of the jet and mass properties as measured at $ r = 10r_g $ in different models.
			{\bf Panel (a)}: Accretion rate on the BH exhibits a continuous decrease over the first second before stabilizing.
			{\bf Panel (b)}: Gravitationally bound mass rapidly decays from $ \sim 0.1\msun $ to $ \sim (10^{-3}-10^{-4})\msun $ as the torus is consumed by the BH.
			{\bf Panel (c)}: When the jet is launched successfully, its launching efficiency ($ \eta = L_{j,0}/\dot{M}c^2 $, where $L_{j,0}$ is the jet power upon launching, or the radial energy flux excluding the rest energy flux) builds over the first $ \sim 0.5 $ s before reaching a MAD state.
			{\bf Panel (d)}: The decrease in the accretion rate leads to luminosity drop to typical sGRB luminosities. If $ \vmin < \vesc $ and $ \beta_p $ is not low enough, massive accretion on the jet axis inhibits a successful jet launching (model $ \alpha $3d1Mb).
		}
		\label{fig:launching}
	\end{figure}
	
	Let us first consider our fiducial model, $\alpha$3d5, shown with thick blue solid lines in Figure~\ref{fig:launching}.  Here and below, unless stated otherwise, we measure all times with respect to the BH formation, when we begin the simulation. Figure~\ref{fig:launching}(a) shows that the massive torus ($ M_t \approx 0.2\msun $) initially drives an accretion rate burst of $ \dot{M} \sim 2\msun\,{\rm s^{-1}} $, such that most of the torus mass is consumed by the BH during the first $\sim 0.1$~s. Consequently, the mass of gravitationally bound gas in Fig.~\ref{fig:launching}(b) rapidly decays to $ M_b \lesssim 10^{-3}\msun $, which results in a further drop in the accretion rate. Several models feature similar accretion rate profiles (e.g., models $ \alpha $3d5, $ \alpha $1d5, $ \alpha $3d5iso and $ V $), however their $ M_b $ time-dependencies are different. The reason is that parts of the ejecta are shocked by the jet and ultimately become bound to increase $ M_b $. When less shocked ejecta is generated, $ M_b $ is lower (e.g. in model $ V $, which does not have the dynamical ejecta). A comparison between models $ \alpha $3d5 and $ \alpha $3d1Mb in Fig.~\ref{fig:launching}(a,b) reveals that if some of the dynamical ejecta is falling onto the BH, i.e. it includes mass with $ v < \vesc $, then the bound ejecta component can dominate $ \dot{M} $ and $ M_b $ already at $ t \sim 0.1 $~s, thereby suppressing the jet launching efficiency early on.
	
	Figure~\ref{fig:launching}(c) shows that accretion of the torus onto the BH leads to the formation of bipolar relativistic jets at $t \sim 20$~ms; by $\sim 0.5$~s, the jet launching efficiency, $\eta = L_{j,0}/\dot Mc^2$, reaches high values, $ \sim 1 $, in all models initialized with poloidal magnetic flux in the torus (except model $\alpha$3d1Mb where the jet fails to emerge as we explain later); here $L_{j,0}$ is the energy carried by highly magnetized regions with $\sigma > 1$ (energy flux minus rest-mass energy flux). This is a tell-tale sign that the disk enters the magnetically arrested disk (MAD; e.g., \citealt{2011MNRAS.418L..79T}) state. Because the accretion rate continues to drop, by the end of the simulation the jet luminosity falls by about an order of magnitude with respect to its value at the onset of the MAD state, as seen in Fig.~\ref{fig:launching}(d). In MADs, the ratio between the fastest growing MRI wavelength mode and the disk scale height scales as $ \beta_p^{-1/2} $ and governs the jet engine activity duration \citep{Masada2007,Gottlieb2022}. For our choice of magnetic flux distribution, we find that when $ \beta_p > 10^3 $, the jet efficiency (and luminosity) drops below unity within $ \lesssim 1 $ s. When the torus is initialized with a toroidal magnetic field, even the magnetic field in equipartition with the gas pressure, the jet power remains very low, consistent with previous simulations of \citet{Christie2019}. However, while \citet{Christie2019} did not consider the dynamical ejecta, we find that such weak jets cannot drill through the dynamical ejecta. This presents challenges to sGRB jet launching from the initially toroidal magnetic field geometry in the presence of polar dynamical ejecta.
	
	Models $ \alpha $3d1Mb and $ \alpha $3d1MbBs stand apart from the other models as they contain $ M_b \approx 0.2\Mej $ at the onset of the simulation. The bound mass leads to a high accretion power of dynamical (unmagnetized) ejecta at the jet launching site. \citet{Pavan2021} showed that this can suppress the jet efficiency and result in an unsuccessful jet launching. For the given isotropic equivalent bound mass along the poles in these models, the minimum jet luminosity required for overcoming the ram pressure of the accreted gas is $ \sim 10^{51}~{\rm erg~s^{-1}} $ \citep{Gottlieb2022}.
	Thus, model $ \alpha $3d1Mb fails to produce a relativistic jet, whereas model $ \alpha $3d1MbBs, which includes strong magnetic fields, enables jet launching, however in this case the jet energy ($ \gtrsim 10^{52} $ erg) is too high to represent an sGRB jet.
	When the ejecta minimal velocity is increased from $ \vmin = 0.05c <\vesc $ to $ \vmin = 0.07c > \vesc $ (model $ \alpha $3d1), unmagnetized ejecta is not falling onto the BH along the poles, and the jets can be successfully launched.
	These results demonstrate that the state of the ejecta at the time of the BH formation is of the utmost importance for the jet's fate. Numerical relativity simulations, which model the jet and ejecta formation self-consistently, show that in fact there is a non-negligible amount of bound mass along the pole at the time of jet formation. Consequently, such simulations feature similarly disrupted jets owing to strong mixing between the jet and the ejecta close to the launching site at $ \sim 10^8 $ cm \citep[][]{Ciolfi2020}.
	
	GN21 previously considered the strength of the jet relative to the surrounding ejecta at the time of jet launching, which dictates the jet propagation. They defined it as
	\begin{equation} 
	    \eta_0 = \left(\frac{L_jt_d}{E_{\rm ej,min}\theta_j^4}\right)^{1/3}~.
	\label{eq:eta0}
	\end{equation}
	Here, $E_{\rm ej,min}$ is the energy of the ejecta at $ \vmin $ and $ \theta_j $ is the jet opening angle. Jets can succeed in drilling out of the dynamical ejecta if the jet engine activity time exceeds the characteristic timescale required for the ejecta to clear out the pole. However, in their analysis, GN21 assumed the jet power to be constant in time, irrespective of the processes that take place at the jet launching site. In reality, the jet power depends on the mass accretion rate, which we show here to be dependent on the inner parts of the ejecta. We find that the magnetized torus is consumed early on, such that the unmagnetized ejecta dominates the accreted gas at later times, thereby suppressing the MAD state after the ejecta expanded. As a result, both the available jet power and efficiency decrease over time until the jet completely shuts off. We conclude that within our torus setup, a successful jet launching requires the magnetic field to be amplified to our prescribed values \citep[e.g. by Kelvin-Helmholtz instability in the merger;][]{Kiuchi2015}. Additionally, there is a minimal delay time between the merger and the BH formation during which the mass along the pole needs to be evacuated before the jets are launched. Failed jets such as in model $ \alpha $3d1Mb cannot to be the source of GRBs, however they may still power low-luminosity emission via cocoon shock breakout \citep{Gottlieb2018b}.
	
	\section{Results}\label{sec:propagation}
	
	In \S\ref{sec:launching}, we showed that some models can produce relativistic jets whereas others fail to do so. Among the models in which the jet fails to pierce through the ejecta, the cocoon can also either break out or not from the dynamical ejecta. Overall, we identify three types of outflow, depending on the value of $ \eta_0 $ (see eq.~\ref{eq:eta0}): (i) {\bf Relativistic Breakout} (RB; high $ \eta_0 $): the jet (and the cocoon) breaks out and powers the prompt $ \gamma $-ray emission for on-axis observers; (ii) {\bf Mildly-relativistic Breakout} (MB; moderate $ \eta_0 $): jets that fail to break out from the ejecta, but their cocoons manage to escape (models $ \alpha $3d1Mb and $ \alpha $3d5Bw); (iii) {\bf Failed Cocoons} (FC; low $ \eta_0 $): both the jet and the cocoon fail to break out from the ejecta because the weak cocoon is moving slower than $ \vmax $. In this section, we focus on the emerging structure from the ejecta and thus ignore failed cocoons (models $ \alpha $3d5Bt and $ \alpha $3d5Bwiso). We note that previously, \citet{Duffell2018} suggested a similar classification of four outflows, three of which are the ones defined above. However, the fourth type, which they defined as a late breakout of a relativistic jet from the ejecta, is attributed to the artificial choice of the jet injection rather than a likely scenario expected in BNSs (see discussion in GN21).
	
	\subsection{Jet propagation}
	
	\begin{figure*}
		\centering
		\includegraphics[scale=0.62,trim=0 0 0 0]{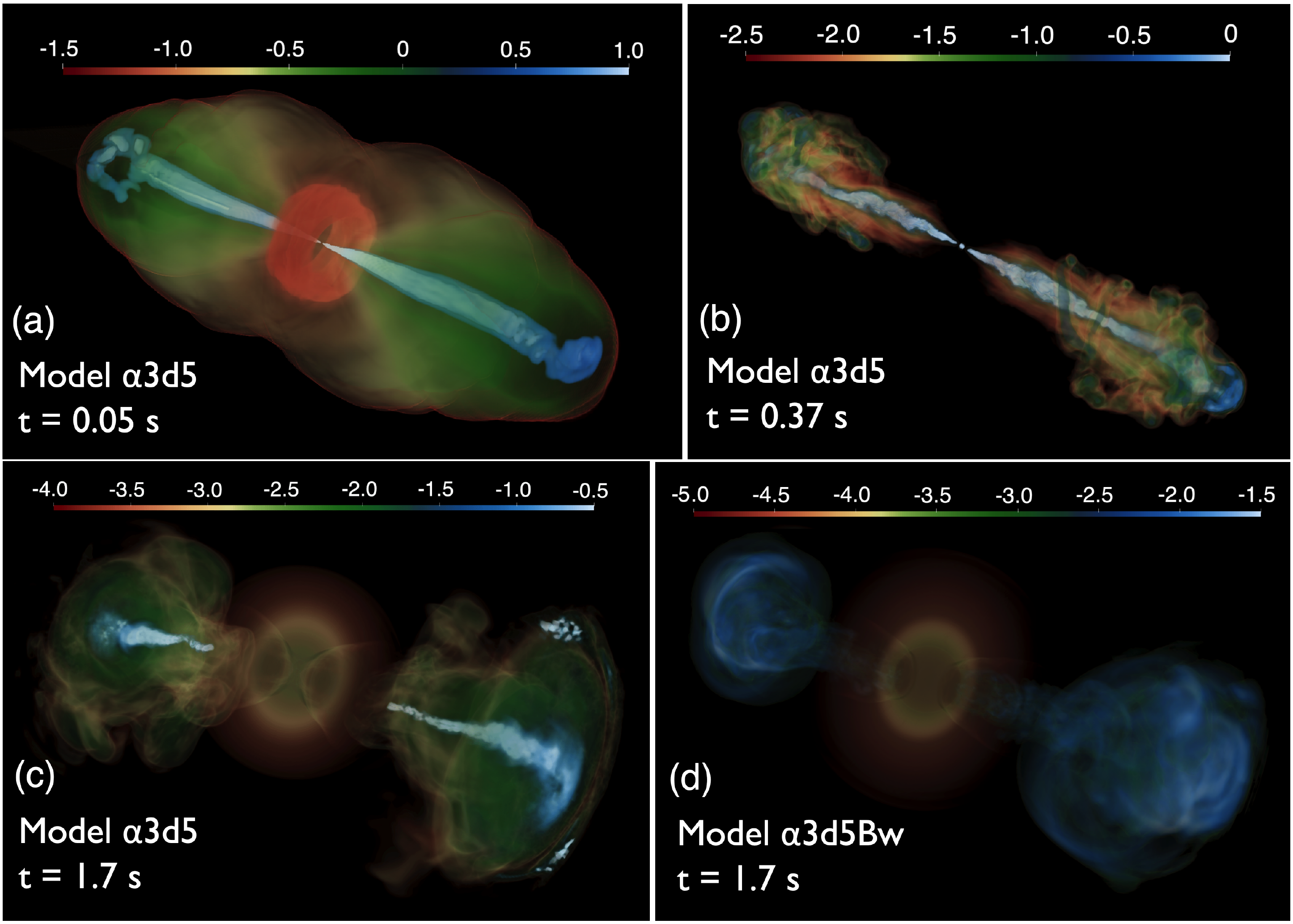}
		
		\caption[]{
			3D rendering of the outflows in different models.
			{\bf Panel (a)}: Logarithm of the asymptotic proper velocity $ u_\infty $ shows the jet (white blue) and cocoon (green) during the interaction with the dense winds (yellow) from the accretion disk (red) at $ r < \vmin t_d $, 0.05 s after the collapse to a BH in model $ \alpha $3d5. Owing to weak mixing between the jet and the winds, the jet retains its high asymptotic proper velocity $ u_\infty $. The qualitative behavior in all models (apart from $ \alpha $3d1Mb in which the jet fails to form) is similar at this stage.
			{\bf Panels (b)} and {\bf (c)}: Logarithm of the magnetization at the time of jet breakout from the ejecta (b) and when the jet head is at $ 4.5\times 10^{10} $~cm (c) in model $ \alpha $3d5. The weak interaction between the jet and the dynamical ejecta does not disrupt the jet which remains intact and highly magnetized.
			{\bf Panel (d)}: Logarithm of the magnetization in the MB model $ \alpha $3d5Bw when the jet head is at $ 4.5\times 10^{10} $~cm. The strong mixing chokes the relativistic gas inside the ejecta, allowing just the weakly magnetized cocoon to break out.
			Panels (v) and (d) also portray the unmagnetized ejecta (which is added to this visualization for completeness) from which the outflows broke out (see movies at \url{http://www.oregottlieb.com/NSM_GRMHD.html}).
		}
		\label{fig:3d}
	\end{figure*}
	
	We find that in all RB models the jet head accelerates to mildly-relativistic velocities (in the ejecta frame) at early times, implying that the jet head is sub-relativistic only during its interaction with the disk winds and the innermost parts of the ejecta. The cocoon energy can be estimated as \citep{Lazzati2005}: $ E_c = L_{j,0}t_b(1-\langle \beta_h\rangle) $, where $ \langle \beta_h \rangle $ is the average dimensionless (normalized by the speed of light) jet head velocity and the $(1-\langle \beta_h\rangle)$ term accounts for the jet material that does not cross the reverse shock by the time the jet head breaks out. This implies that the jet head primarily energizes the cocoon while the head is sub-relativistic. Because the head is sub-relativistic only during the first few dozen milliseconds -- while interacting with the dense winds, -- this interaction plays the most significant role in determining the complex jet structure. It then follows that if the jet is choked inside the ejecta, this can only happen very early on in jet-ejecta interaction. We show in \S\ref{sec:structure} that RB models feature similar post-breakout outflow structure and properties regardless of the presence of the dynamical ejecta. Indeed, the dynamical ejecta appears to make a difference in the post-breakout outflow structure only for FC and MB models.
	
	Figure~\ref{fig:3d} shows a 3D rendering of the system at different times and models. Fig.~\ref{fig:3d}(a) depicts the logarithm of the asymptotic proper velocity, $ u_\infty $ at early times (before its interaction with the ejecta at $ r = \vmin t_d $) in model $ \alpha $3d5. Here, $ u_\infty \equiv (\Gamma_\infty^2-1)^{1/2} $, where $ \Gamma_\infty = -u_t\left(h+\sigma\right) $ is the asymptotic Lorentz factor, $ u_t $ is the covariant time-component of the four-velocity, and $ h \equiv 1+4p_g/\rho c^2$ is the specific enthalpy. It is shown that the early interaction between the jet (white blue) and the winds (yellow) from the disk (red) is strong enough to form strong shocks at the jet head that lead to an extended structure of a hot cocoon (green) prior to the its interaction with the ejecta at $ \vmin(t+t_d) $. The remaining panels of Fig. \ref{fig:3d} portray the logarithm of the magnetization in the jet and the cocoon, upon breakout from the dynamical ejecta (right panel, model $ \alpha $3d5), and when the forward shock is at $ r = 4.5\times 10^{10} $ cm (Fig.~\ref{fig:3d}(c) for RB model $ \alpha $3d5 and Fig.~\ref{fig:3d}(d) for MB model $ \alpha $3d5Bw). Figs.~\ref{fig:3d}(b) and (c) feature a jet that remains intact and retains a high magnetization\footnote{In order for the jet to reach observed Lorentz factors of a few hundreds, we expect that $ \sigma_0 \gg 20 $. In the absence of mixing, we expect that the magnetization in the jet maintains a similar profile, but with a higher normalization, proportional to the adopted value of $ \sigma_0 $.} of $ \sigma \sim 1 $ both before and after the breakout, whereas the cocoon magnetization is $ \sigma < 10^{-2} $. The weak jet model in Fig.~\ref{fig:3d}(d) features a post-breakout cocoon without a jet. Here, the weak jet was choked inside the ejecta, implying that jets with $ L_j \approx 10^{49}~{\rm erg\,s^{-1}} $ can only emerge from a lighter ejecta. We test this hypothesis by performing a simulation corresponding to model $ V $Bw (Table~\ref{tab:models}), for which we repeat simulation $ \alpha $3d5Bw, but remove the dynamical ejecta. We find that the weak jet can successfully propagate through the disk winds and form a similar structure to that in model $ \alpha $3d5. We conclude that while the ejecta has a negligible effect on jets that break out of it, it sets a lower limit on the energy of jets that can escape from it.
	
	\begin{figure}
		\centering
		\includegraphics[scale=0.19,trim=0 0 0 0]{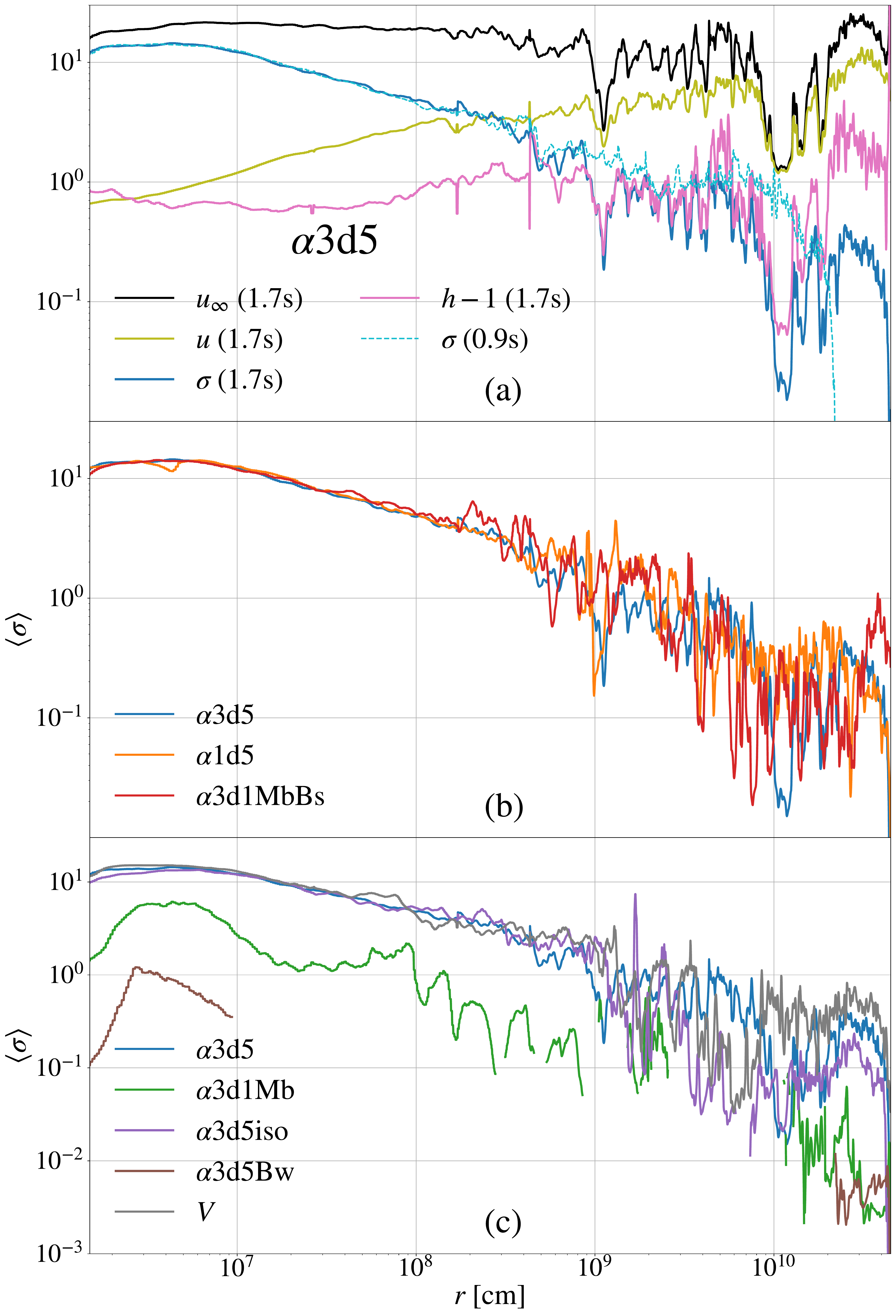}
		\caption[]{
			Radial profiles of radial energy flux weighted averages, excluding the rest mass energy flux and considering only matter with $ u_\infty > 1 $.
			{\bf Panel (a)}: Efficient conversion of magnetic energy into kinetic during the early jet launching in model $ \alpha $3d5, before the jet power becomes too weak to retain its initial energy (inner $ \sim 2\times 10^{10} $ cm).
			Shown are two different times: when the jet head reaches $ 4.5 \times 10^{10} $ cm $ 1.7 $ s after the merger (solid thick lines) and $ 0.9 $ s after the merger (dashed thin line).
			{\bf Panel (b)}: Similar magnetization profiles in models $ \alpha $3d5, $ \alpha $1d5 and $ \alpha $3d5Bs when the jet head is at $ r = 4.5\times 10^{10} $ cm, implying that the magnetization profile is almost independent of the specific properties of the system as long as the jet successfully breaks out from the ejecta.
			{\bf Panel (c)}: Magnetization profiles of models with different jet power or polar densities when the jet head is at $ r = 4.5\times 10^{10} $ cm demonstrate that low $ \eta_0 $ results in stronger dissipation. The highest average magnetization is obtained without initial ejecta (model $ V $), stronger dissipation is observed when ejecta is falling toward the launching point ($ \alpha $3d1Mb) or when the jet power is low ($ \alpha $3d5Bw).
		}
		\label{fig:sigma}
	\end{figure}
	
	Figure~\ref{fig:sigma} depicts the radial energy flux (excluding the rest mass energy flux) weighted average of different quantities and models, while considering only material with $ u_\infty > 1 $. Fig.~\ref{fig:sigma}(a) demonstrates that during the first $ \sim 1 $ s, the magnetization profile in model $ \alpha $3d5 decays monotonically as $ r^{-1/2} $ (dashed light blue line) until it reaches the head. The magnetization profile at 1.7 s (blue) shows that it is efficiently converted to bulk kinetic energy (olive) such that the maximum velocity of a fluid element $ u_\infty $ (black) is conserved, implying that jet-cocoon mixing is insignificant. Both the magnetization and the specific enthalpy (magenta) remain moderate in the jet at $ \sim 10^{10.5}$~cm. While the jet power is getting weaker, the structure becomes more intermittent and the mixing is getting stronger \citep{Gottlieb2020a}, as seen in the drop at $ \sim 10^{10} $ cm in the profiles at $ 1.7 $~s.
	
	Figure~\ref{fig:sigma}(b) demonstrates that a similar radial profile of the magnetization is also observed in all other models in which the jet encounters a similar density profile along its path in the ejecta. Figure~\ref{fig:sigma}(c) reveals subtle differences between RB models -- when the expanding ejecta is absent (model $ V $), the mixing is somewhat weaker, and at the largest radii the magnetization remains slightly above the models with the dynamical ejecta. When the ejecta is isotropic (model $ \alpha $3d5iso), more mass is present on the jet axis, the mixing is stronger and the magnetization is slightly lower. Substantial differences appear between the magnetization profile of RB and MB models. The drop in the jet energy over time in model $ \alpha $3d5Bw increases the mixing between the jet and the ejecta, such that the magnetization profile is truncated at $ 10^7 $~cm as no relativistic material is present. The only relativistic material with $ u_\infty > 1 $ is observed in the jet that was launched at early times and is seen at $ \sim 3 \times 10^{10} $ cm, where $ \sigma \sim 10^{-2} $. Similarly, when the jet is not successfully launched (model $ \alpha $3d1Mb), the wide outflow from the BH is subject to strong mixing which also results in a weak magnetization of the outflow. Overall, when $ \eta_0 $ is low, the jet-ejecta interaction is strong and the magnetization of the jet decreases well below unity. In contrast, in typical GRB jets (RB models) $ \eta_0 $ is high, such that the jet head propagates relativistically and the jet maintains $ 0.1 \lesssim \sigma \lesssim 1 $ (see Fig.~\ref{fig:Edist}).
	
	\subsection{Jet structure and emission}\label{sec:structure}
	
	\begin{figure*}
		\centering
		\includegraphics[scale=0.2,trim=0 0 0 0]{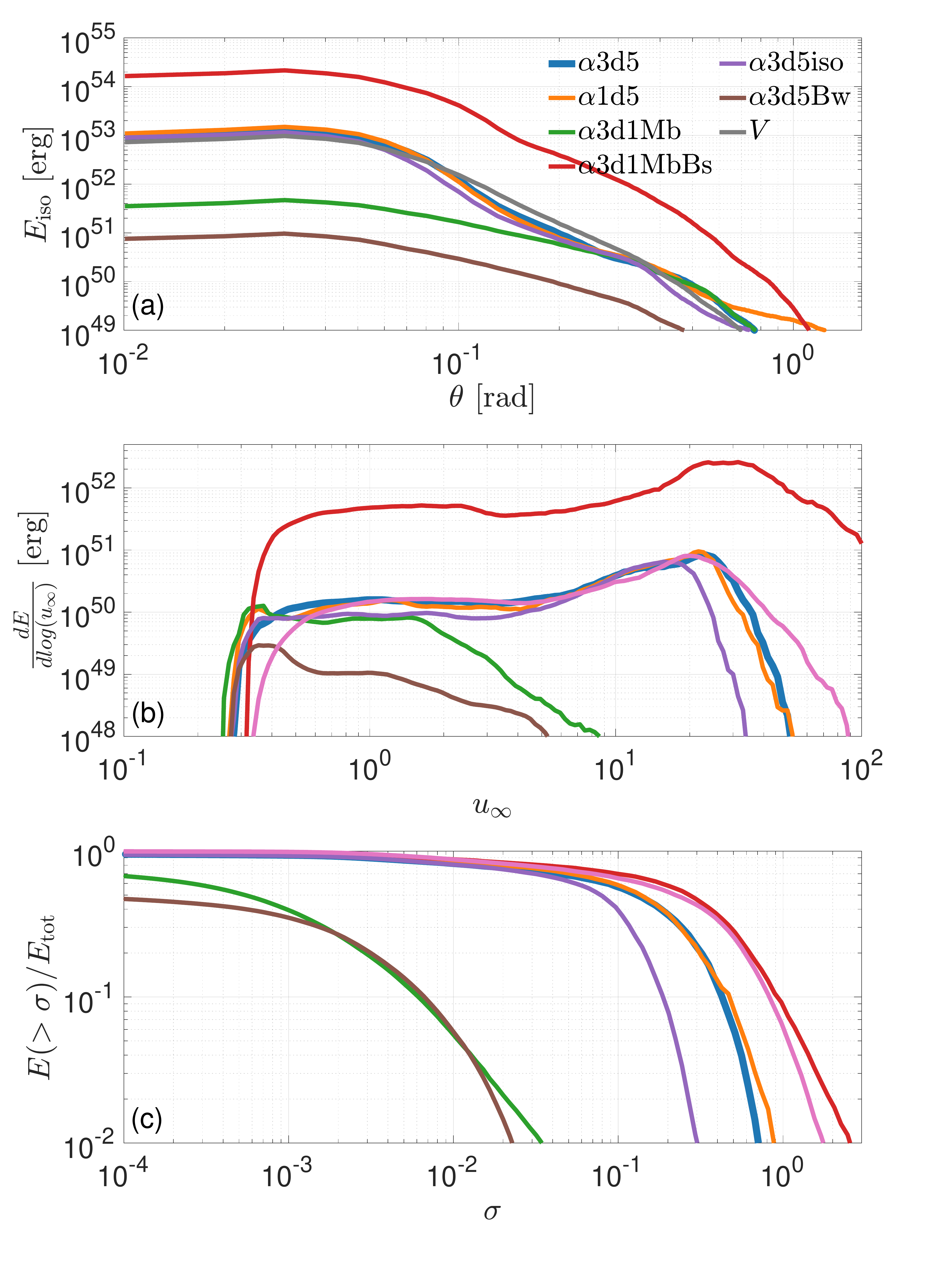}
		\includegraphics[scale=0.2,trim=0 0 0 0]{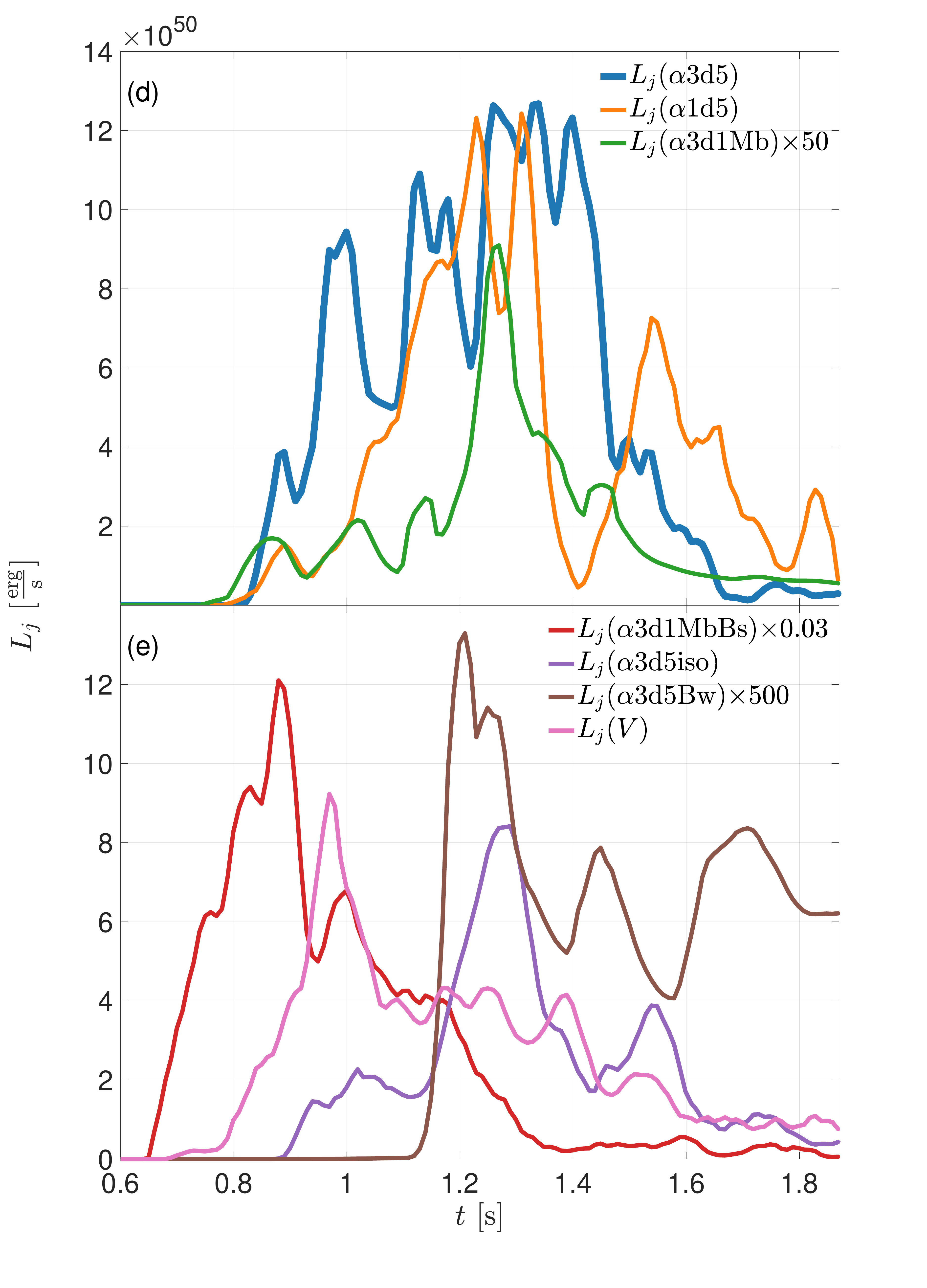}
		\caption[]{
			{\bf Panels (a)--(c):} Profiles of the different models when the jet head is at $ 4.5\times 10^{10} $ cm.
			{\bf Panel (a)}: Angular profiles of the isotropic equivalent energy of the material outside the ejecta when the jet head reaches $ 4.5\times 10^{10} $ cm hint at a universal structure with a flat core followed by a power-law. The ejecta is important only when the jet fails to break through it.
			{\bf Panel (b)}: The energy distribution per logarithmic asymptotic proper velocity of the material that broke out from the ejecta. A uniform distribution is shown in the cocoon between $ u_\infty \approx \vmax $ and $ u_\infty \sim 3 $. The low level of mixing allows the jet to retain its energy such that the energy rises in ultra-relativistic velocities.
			Both angular and radial distributions are consistent with those of hydrodynamic jets \citep{Gottlieb2021a}.
			{\bf (panel (c)}: The cumulative energy in gas with magnetization larger than $\sigma$ out of the total energy of the gas outside the ejecta. When a relativistic jet is present, most of the plasma maintains $ \sigma > 0.1 $, which is expected to be even higher when a realistic $ \sigma_0 > 100 $ is used.
			{\bf Panels (d)} and {\bf (e)}: The total jet power post-breakout (integrated over a shell of radius $ 2\times 10^{10} $~cm). It shows a highly variable $\sim 0.5$~s long signal whose energetic varies from model to model (with a total energy spanning the range of $10^{48}{-}10^{52}$~erg).
		}
		\label{fig:Edist}
	\end{figure*}
	
	Fig. \ref{fig:Edist}(a) depicts the angular structure of the jet-cocoon outflow by integrating in the radial and azimuthal directions over all the material that broke out of the ejecta. All models exhibit the jet structure with a flat core and power-law decay $ E_{\rm iso}(\theta) \propto \theta^{-\delta} $.  Interestingly, a similar structure was previously found by \citet{Gottlieb2021a,Nativi2022} and \citet{Gottlieb2020b} in hydrodynamic and weakly magnetized jets, respectively. Short GRB simulations that do not last long enough for the jet to build its asymptotic extended structure feature a different angular structure \citet{Kathirgamaraju2019,Nathanail2021}. Because $ \langle \beta_h \rangle $ is mildly-relativistic in RB models, the jet structure is governed by the interaction with the winds and therefore it is only weakly dependent on the inclusion of the ejecta. In these models the power-law index is $ \delta \gtrsim 3 $ due to a low level of mixing, and consistent with the power-law indices found for hydrodynamic and weakly magnetized sGRB jets \citep{Gottlieb2020b,Gottlieb2021a}. In the MB models ($ \alpha $3d1Mb, $ \alpha $3d5Bw), only the mildly-relativistic is present outside the ejecta, and thus the expansion is to wider angles. Consequently, the outflow has a milder power-law index $ \delta \sim -1.5 $ (green and brown lines), similar to long GRBs \citep[lGRBs;][]{Gottlieb2021a} where the jet energy/ejecta density ratio is similarly lower.
	
	Figure~\ref{fig:Edist}(b) shows the distribution of the energy outside the ejecta in the logarithm of $u_\infty$. In the range of cocoon velocities, $ \vmax/c \lesssim u_\infty \lesssim 3 $, the distribution is quasi-uniform in all models, in agreement with previous results of hydrodynamic and weakly magnetized jets \citep{Gottlieb2020b,Gottlieb2021a}. In RB models, the transition to ultra relativistic velocities shows a rise in the energy until $ u_\infty \sim \sigma_0 $, owing to the low mixing of the jet with the ejecta.
	Figure~\ref{fig:Edist}(c) shows the fraction of energy above a certain value of magnetization. Most of the jet energy in the RB models is carried by plasma with $ \sigma > 0.1 $. Under the reasonable assumption that a higher $ \sigma_0 $ does not increase the mixing between the jet and the ejecta, it follows that for $ \sigma_0 \gtrsim 100 $ (as expected in GRBs), most of the jet energy would come out at $ \sigma \sim 1 $. This result demonstrates that magnetic energy has an essential contribution to the powering signal of typical sGRBs, e.g. via synchrotron emission and/or magnetic reconnection. Conversely, we see that in the absence of the jet in the MB models, the magnetization of the outflow is negligible (green and brown curves).
	
	Figures~\ref{fig:Edist}(d) and (e) portray the total power (calculated in the same manner as the jet power on the horizon $ L_{j,0} $) at $ r = 2\times 10^{10} $ cm, after the breakout from the ejecta. The high variability over a $ \sim 10-100 $ ms timescale with $ \sim 50\% $ power fluctuations and the $ \sim 0.5 $ s duration of the signal, resemble observational features of sGRB lightcurves. In RB models $ \alpha $3d5, $ \alpha $1d5, $ \alpha $3d5iso and $ V $, the total energy of $ \sim 10^{50.5} $ erg is also consistent with observations. The energies of the MB models $ \alpha $3d1Mb and $ \alpha $3d5Bw, $ 1-5\times 10^{48} $ erg, may produce low-luminosity sGRBs.
	We emphasize that GRB lightcurves could be considerably different from the jet power that we show here, because the emission depends on the radiative efficiency and composition of the jet. It is notable that both the magnetization and the specific enthalpy are high (Fig.~\ref{fig:sigma}(a)), such that both components contribute to the emission. We plan to calculate the jet and cocoon emission based on our simulations in future work.
	
	\section{Conclusions}\label{sec:conclusions}
	
	We presented the first 3D GRMHD sGRB jet simulations from the BH to $ \sim 3\times 10^{10} $ cm with the highest resolution of such simulations to date. There are three types of outflows from the merger: sub-relativistic ejecta, mildly-relativistic cocoon and relativistic jets. We find that there are three possible outcomes, which depend on $ \eta_0 $ -- the ratio of jet to ejecta energy at the time of jet launching: (i) Relativistic breakout (high $ \eta_0 $): all components are present in the coasting phase; (ii) Mildly-relativistic breakout (moderate $ \eta_0 $): the jet fails to emerge from the ejecta; (iii) Failed cocoon (low $ \eta_0 $): the jet and cocoon are both choked inside the ejecta. Given the large data set of sGRB jets, their luminosity range can be well estimated \citep[e.g.,][]{Wanderman2015}. Although there are only a handful of kilonova observations from which the ejecta mass can be inferred, they hint at a few $ 10^{-2} \msun $ ejecta \citep{Tanvir2013,Berger2013,Kasen2017,Rastinejad2022}. We note that while $ \Mej $ is composed of dynamical ejecta and disk winds, the important mass for the jet interaction is along the pole, and is expected to be dominated by the dynamical ejecta component. For example, in sGRB 160821B the inferred total ejecta mass was $ \Mej \sim 0.01 \msun $ but only $ \sim 10^{-3}\msun $ along the pole, allowing the relatively weak ($ \sim 10^{49} $ erg) jet to break out \citep{Lamb2019}. However it is possible that these massive ejecta are subject to observational bias. Using the above estimates, we considered different configurations of the torus, ejecta and delay times to characterize the dependency of the jet properties and its emerging post-breakout structure on the underlying physics of the merger. Our main conclusions are as follows:
	
	\begin{itemize}
		
		\item {
		    We can constrain the ejecta mass and the time delay between the merger and the BH formation, based on the success of launching a relativistic jet.
		    If bound mass along the poles still exists at the time of jet formation, the jet may fail to proceed if it cannot overcome the ram pressure of the infalling ejecta. In contrast to previous studies in which the jet power was a free parameter chosen as part of the setup, we find that the accretion onto the BH becomes dominated by the unmagnetized ejecta after a short while. This leads to a drop in the jet efficiency such that a relativistic jet cannot be launched. Therefore, there is a minimal time delay after the merger during which the bound ejecta isotropic equivalent mass along the pole has to drop (roughly below $ \sim 10^{-4}\msun $ for typical sGRB jets) before the BH forms and launches the relativistic jets. The minimal jet power needed for the jet to break out of the ejecta with an isotropic equivalent mass of $ \sim 10^{-2} \msun $ along the pole is $ L_j \gtrsim 10^{50}~{\rm erg~s^{-1}} $. This suggests that the low luminosity end of sGRB distribution requires a lower polar ejecta mass in order for the jet to break out, in agreement with \citet[][]{Gottlieb2021a} and \citet{Gottlieb2021d}.
		}
		\item {
		    In order for the jet to be successfully launched, the magnetic field amplification should take place either before, or $ \lesssim 0.1 $ s after, the BH formation, before most of the disk is accreted.
		}
		\item {
			We find that jets launched in models featuring BH disks with the toroidal field geometry are too weak to propagate through the dynamical ejecta at any time due to the above arguments (even for extremely strong magnetic fields, $ \beta_p = 1 $). This result disfavors physically motivated BH disks with a toroidal magnetic field configuration as the source of sGRB jets. This implies that the magnetic field might be altered prior to the BH formation, or that neutrino physics is important for generating poloidal fields in the disk. The latter will be addressed in a future work.
		}
		\item{
		    Within our framework, we find that the torus plasma beta needs to have $ \beta_p \lesssim 10^3 $, or the magnetic field strength needs to be $ B \gtrsim 10^{15} $~G, in order to successfully launch sGRB jets.
		}
		\item {
			The post-breakout angular distribution of the isotropic equivalent energy of the outflow is a universal structure, irrespective of the jet magnetization, with a flat core followed by a power-law decay. If the jet is strong enough to break out from the ejecta, then its head is at least mildly-relativistic and the dynamical ejecta effect on the jet is negligible. The power-law index in this case is $ \gtrsim 3 $, consistent with hydrodynamic jets \citep{Gottlieb2021a}. If the jet is suffocated in the ejecta and only the cocoon emerges, then the power-law index is $ \sim 1.5 $.
		}
		\item {
			The BH disk ejects quasi-isotropic baryon loaded winds, which ultimately shape the jet structure, such that the further jet-ejecta interaction does not change the jet properties to a large extent. This implies that the ejecta is not essential for obtaining an extended angular jet structure: the jet may gain its complex structure solely through the interaction with the winds from the BH disk. Further investigation of the wind properties is left for a future study that includes a neutrino scheme.
		}
		\item{
			In a companion paper, \citet{Gottlieb2022b}, we study the evolution of lGRB jets in collapsars. Qualitatively, the main difference between the problems is the jet energy/ejecta density ratio. In collapsars, the medium is more massive, such that the above ratio is lower, similar to our models with weak jets. As a result, the lGRB jet head propagates slower inside the medium. In contrast to the dynamical ejecta, stars likely have a structured magnetic field, and thus the jet can ultimately break out thanks to the accretion of the mass reservoir in the star that allows the jet central engine to operate over much longer timescales. Here, we find that after breaking out from the medium, a typical sGRB jet remains intact whereas in collapsars the interaction with the massive star leads to an intermittent structure. Similarly, the post-breakout magnetization level is somewhat higher in sGRB jets with $ \sigma \gtrsim 0.1 $ (and potentially higher for more realistic $ \sigma_0 \gtrsim 100 $). \citet{Gottlieb2022b} showed that after the jet escapes from the dense medium, the magnetization remains unchanged and the jet reaches the photosphere with the same magnetization with which it broke out. This implies that the magnetization found here at $ \sim 10^{10.5} $ cm would be similar to that at the photosphere at $ \sim 10^{12} $ cm, implying that the magnetic energy may well contribute to shaping the non-thermal spectrum of the sGRB prompt emission. The strong jet-medium interaction in collapsars also results in disk tilt and jet wobble that increases the angle for detection, thereby decreasing the intrinsic GRB rate, and also produces quiescent times in the lightcurve. While our simulations do not feature a wobbling jet, we attribute this result to our choice of initial conditions that include a stable torus, rather than an intrinsic difference between sGRB and lGRBs. It is likely that for a self-consistent disk formation a similar behavior will emerge.
		}			
	\end{itemize}
	
	\begin{acknowledgements}
		
	We thank Om Sharan Salafia for useful comments.
	OG is supported by a CIERA Postdoctoral Fellowship.
    OG and AT acknowledge support by Fermi Cycle 14 Guest Investigator program 80NSSC22K0031.
    AT was supported by NSF grants
    AST-2107839, 
    AST-1815304, 
    AST-1911080, 
    AST-2031997, 
    and NASA grant 80NSSC18K0565.
    AT was partly supported by an NSF-BSF grant 2020747.
    AMB is supported by NASA through the NASA Hubble Fellowship grant HST-HF2-51487.001-A awarded by the Space Telescope Science Institute, which is operated by the Association of Universities for Research in Astronomy, Inc., for NASA, under contract NAS5-26555 and NASA award TCAN-80NSSC18K1488.
    An award of computer time was provided by the Innovative and Novel Computational Impact on Theory and Experiment (INCITE) program under award PHY129. This research used resources of the Oak Ridge Leadership Computing Facility, which is a DOE Office of Science User Facility supported under contract DE-AC05- 00OR22725.
    This research used resources of the National Energy Research Scientific Computing Center, a DOE Office of Science User Facility supported by the Office of Science of the U.S. Department of Energy under Contract No. DE-AC02-05CH11231 using NERSC award NP-ERCAP0020543 (allocation m2401).
    The authors acknowledge the Texas Advanced Computing Center (TACC) at The University of Texas at Austin for providing HPC and visualization resources that have contributed to the research results reported within this paper via LRAC allocation AST20011 (\url{http://www.tacc.utexas.edu}).
    This research was also enabled in part by support provided by Compute Canada allocation xsp-772 (\url{http://www.computecanada.ca}).
		
	\end{acknowledgements}
	
	\section*{Data Availability}
	
	The data underlying this article will be shared on reasonable request to the corresponding author.

	\bibliography{refs} 
	
\end{document}